\documentclass[aps,iop,twocolumn,amsmath,longbibliography, amssymb,superscriptaddress]{revtex4-1}

\pdfoutput=1
\usepackage{bm}
\usepackage[retainorgcmds]{IEEEtrantools}
\usepackage{graphicx}
\usepackage{mathrsfs}
\usepackage{amsmath}
\usepackage{amssymb}
\usepackage{color}
\usepackage{amsfonts}
\usepackage{times,txfonts}
\usepackage{nicefrac}
\usepackage[colorlinks=true,linkcolor=blue,urlcolor=blue,citecolor=blue,pdfusetitle]{hyperref}
\usepackage{soul}
\usepackage[dvipsnames]{xcolor}

\newcommand{\belfa}[1]{\textcolor{black}{#1}}
\usepackage{verbatim}
\usepackage{placeins}
\usepackage{tikz}
\usetikzlibrary{decorations.pathmorphing}
\newcommand{\ket}[1]{\vert#1\rangle}
\newcommand{\abs}[1]{\vert#1\vert}
\newcommand{\op}[2]{\vert#1\rangle\langle#2\vert}
\newcommand{\comm}[2]{\left[#1,#2\right]}

\begin{document}

\title{Reinforcement learning-enhanced protocols for coherent population-transfer in three-level quantum systems}
\date{\today}
\author{Jonathon Brown}\thanks{These authors contributed equally to this work}
\affiliation{Centre for Theoretical Atomic, Molecular, and Optical Physics, School of Mathematics and Physics, Queens University, Belfast BT7 1NN, United Kingdom}
\author{Pierpaolo Sgroi}\thanks{These authors contributed equally to this work}
\affiliation{Centre for Theoretical Atomic, Molecular, and Optical Physics, School of Mathematics and Physics, Queens University, Belfast BT7 1NN, United Kingdom}
\author{Luigi Giannelli} 
\affiliation{Dipartimento di Fisica e Astronomia ``Ettore Majorana", Universita' di Catania, Via S. Sofia 64, 95123, Catania, Italy}
\affiliation{INFN, Sez. Catania, 95123, Catania, Italy}
\author{Gheorghe Sorin Paraoanu}
\affiliation{QTF Centre of Excellence, Department of Applied Physics,
Aalto University School of Science, P.O. Box 15100, FI-00076 AALTO, Finland}
\author{Elisabetta Paladino}
\affiliation{Dipartimento di Fisica e Astronomia ``Ettore Majorana", Universita' di Catania, Via S. Sofia 64, 95123, Catania, Italy}
\affiliation{INFN, Sez. Catania, 95123, Catania, Italy}
\affiliation{CNR-IMM, UoS Universita', 95123, Catania, Italy} 
\author{Giuseppe Falci} 
\affiliation{Dipartimento di Fisica e Astronomia ``Ettore Majorana", Universita' di Catania, Via S. Sofia 64, 95123, Catania, Italy}
\affiliation{INFN, Sez. Catania, 95123, Catania, Italy}
\affiliation{CNR-IMM, UoS Universita', 95123, Catania, Italy} 
\author{Mauro Paternostro}
\affiliation{Centre for Theoretical Atomic, Molecular, and Optical Physics, School of Mathematics and Physics, Queens University, Belfast BT7 1NN, United Kingdom}
\author{Alessandro Ferraro}
\affiliation{Centre for Theoretical Atomic, Molecular, and Optical Physics, School of Mathematics and Physics, Queens University, Belfast BT7 1NN, United Kingdom}

\begin{abstract}
We deploy a combination of reinforcement learning-based approaches and more traditional optimization techniques to identify optimal protocols for population transfer in a multi-level system. We constraint our strategy to the case of fixed coupling rates but time-varying detunings, a situation that would simplify considerably the implementation of population transfer in relevant experimental platforms, such as semiconducting and superconducting ones. Our approach is able to explore the space of possible control protocols to reveal the existence of efficient protocols that, remarkably, differ from (and can be superior to) standard Raman, STIRAP or other adiabatic schemes. The new protocols that we identify are robust against both energy losses and dephasing. 
\end{abstract}

\maketitle 

\section{Introduction}
It is well known that quantum systems can provide clear computational advantage when compared with their classical counterparts, and several algorithms have been presented whereby this advantage is exploited to carry out so called super-classical tasks~\cite{nielsen2000quantum, shor1999polynomial, grover1996fast}. The required control over quantum systems, however, still remains the biggest challenge for full implementation of quantum computing algorithms. An experimental platform that provides a promising candidate for controlling general quantum systems are superconducting circuits, which have been widely employed to fabricate qubits (see~\cite{kjaergaard2020superconducting} and \cite{devoret2013} for reviews) and two qubit gates~\cite{dicarlo2009demonstration,mckay2016universal, caldwell2018parametrically}, as well as implementations of so-called circuit-quantum electrodynamics (QED)~\cite{blais2007quantum, blais2020circuit}, which is at the forefront of the current ``quantum race"~\cite{google}. Multi-level dynamics has also been addressed both theoretically~\cite{falci2009,falci2017advances,distefano2015population,11-nori} and experimentally~\cite{paraoanu2019superadiabatic, kumar2016stimulated,siddiqi2021}. However, together with the promise of an experimental platform to manipulate quantum systems towards the achievement of a quantum network of multiple nodes, comes an increased demand for quantum-control schemes pertinent to the experimental constraints at play. Much of the work towards this goal employs techniques from NMR, Quantum Optics and Quantum Optimal Control theory~\cite{dalessandro2007introduction,15-glaser}. \belfa{Specifically, gradient-based optimization methods have been recently employed to control general open systems with a myriad of applications~\cite{abdelhafez2019gradient}, as well as aiding the design of high-fidelity, protected superconducting quantum gates~\cite{abdelhafez2020universal,xu2020nonadiabatic, werninghaus2021leakage, propson2021robust}.  In} the context of multi-level systems, the use of two tone pulses allow faithful, selective and robust single-qubit quantum operations such as population transfer and generation of superposition and can be generalized to quantum operations on multiple nodes of a network, such as state shuttling, entanglement and single photon generation. Stimulated Raman Adiabatic Passage (STIRAP) and Raman oscillations are two well-known protocols for the implementation of these quantum operations. Some effort has been made to adapt the original formulation of such protocols to reduced-control architectures~\cite{distefano2015population,falci2017advances,16-distefano} or to improve them by using optimal pulse shaping and superadiabatic techniques~\cite{ vasilev2009optimum,paraoanu2019superadiabatic,giannelli2014}. 

More recently machine learning techniques have emerged as a viable option for finding alternative optimal control schemes. \belfa{In particular reinforcement learning (RL) has been employed in the context of state preparation~\cite{sivak2021model,haug2020classifying}, circuit architecture design~\cite{kuo2021quantum} and control of multi-level systems~\cite{an2021quantum}. In the context of three level systems, deep neural network based RL has been used} along with state monitoring to learn optimal pulse shapes for driving fields~\cite{paparelle2020digitally, porotti2019coherent}. Here we implement a two-step optimisation approach, that combines different optimization approaches. Initially, Deep Reinforcement Learning (DRL)-like techniques, in conjunction with Recurrent Neural Networks (RNNs), is used to learn the shape of efficient piece-wise constant control pulses, without the requirement for state monitoring~\cite{sgroi2020reinforcement}. Such key insight is then used to implement a suitable traditional optimization method. This two-step approach yields smooth, analytically well-defined control pulses. {An important point to make is that application of such conventional optimization methods without any pre-available information is much more difficult in general due to the ``curse of dimensionality''~\cite{curse}. To succeed, they require the choice of a suitably truncated basis upon which to expand their control functions. This highlights the utility of the initial learning step, which is essentially user-independent and can provide a suitable ansatz without the need for prior knowledge of the system. For example, a requirement for the success of STIRAP is the existence of (a manifold of) adiabatic dark states, and the full knowledge of their structure~\cite{vitanov2017stimulated}. On the other hand, for Raman oscillations, the hallmark for adiabatic elimination is the validity of restrictive parameter conditions (such as large detunings), so as to constrain the dynamics to relevant subspaces. The RL-based step discussed here provides protocols that violate both such restrictive conditions, and thus differ from both STIRAP and simple adiabatic elimination, while combining advantages of both to achieve near-optimal dynamics. This thus provides an ansatz for the control that may otherwise not have been arrived at analytically, and whose flexibility could be exploited to engineer operations in multi-node architectures. While delivering previously unforeseen protocols, this hybrid approach to optimisation marks a significant departures form previous methods towards the control of quantum dynamics, and embodies one of the pillars of our proposal.}

The remainder of this paper is organized as follows. In Sec.~\ref{Prelude} we introduce the physical system of interest, which allows us to motivate the specific form of control chosen. In Sec.~\ref{ReinforcementLearning} we show how an RL agent was able to learn control schemes to induce some desired dynamics in the system. Then, in Sec.~\ref{CoeffOptimization}, we use a less sophisticated coefficient optimization over a polynomial basis in an attempt to reproduce the results obtained by the RL approach. In Sec.~\ref{OptimalProtocols} we use the results from the RL agent in Sec.~\ref{ReinforcementLearning}, followed by the simpler coefficient optimization, where we were able to obtain further improvement in protocol efficiency when compared with both methods alone. We then dedicate Sec.~\ref{Decay} to an analysis of the resilience of the learned protocols to stochastic decay within the system, where we explicitly consider the performance of both protocols in a 3-level Ladder system. We finally discuss the robustness of the protocol to low-frequency noise and its resilience to pure dephasing in the system dynamics in Sec.~\ref{Dephasing}, followed by a brief discussion of the results in Sec.~\ref{conc}.

\section{The system }
\label{Prelude}
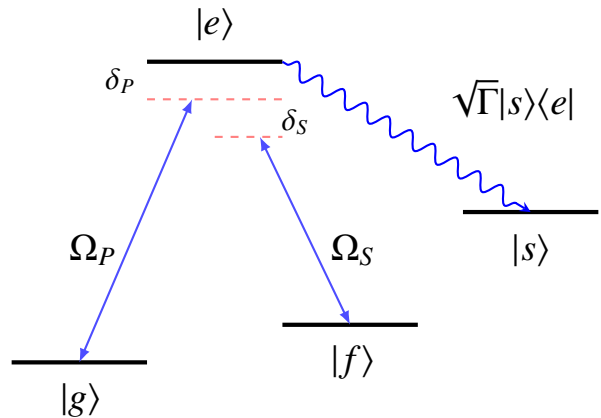
\begin{figure}[t]
\centering
\begin{minipage}{0.43\textwidth}
	\begin{tikzpicture}[x=1.2cm, y=1cm]
		\draw[color=black, ultra thick] (1.5,4)--(3,4);
		\node[text width = 0.5cm] at (2.25,4.5) {\Large $\ket{e}$};
		\draw[color=black, ultra thick] (5,2)--(6.5,2);
		\node[text width = 0.5cm] at (5.75,1.5) {\Large $\ket{s}$};
		\node[text width = 0.5cm, color=black] at (5,3.5) {\Large $\sqrt{\Gamma} \op{s}{e}$};
		\draw[color=black, ultra thick] (0,0)--(1.5,0);
		\node[text width = 0.5cm] at (0.75, -0.5) {\Large $\ket{g}$};
		\draw[color=black, ultra thick] (3,0.5)--(4.5,0.5);
		\node[text width = 0.5cm] at (3.75,0) {\Large $\ket{f}$};
		\draw[color=blue!70, thick, latex-latex] (0.75,0) -- (2,3.5);
		\node[text width = 0.5cm] at (0.85,1.5) {\Large $\Omega_P$};
		\draw[color=red!50, dashed, thick] (1.5,3.5)--(3,3.5);
		\node[text width = 0.5cm] at (1.25,3.75) {\large $\delta_P$};
		\draw[color=blue!70, thick, latex-latex] (3.75,0.5)--(2.75,3);
		\node[text width = 0.5cm] at (3.75,1.5) {\Large $\Omega_S$};
		\draw[color=red!50, dashed, thick] (2.25,3)--(3,3);
		\node[text width = 0.5cm] at (3.2,3.2) {\large $\delta_S$};
		\draw[-stealth,color=blue, decorate,decoration={snake,amplitude=3pt,pre length=2pt,post length=3pt}, thick] (3,4) -- (5.75,2);		
	\end{tikzpicture}
\end{minipage}
\caption{Schematic energy level structure for the Lambda system modulated by detuned AC-driving fields with respective Rabi frequency $\Omega_P$ and $\Omega_S$. The scheme includes a sink state $\ket{s}$, coupled to the excited state $\ket{e}$ of the system, as a means to induce a loss mechanism at rate $\Gamma$.}
\label{fig1}
\end{figure}

We investigate control protocols for an abstract 3-level quantum systems and specifically consider the task of population transfer in so-called {Lambda} systems, where a ground state $\ket{g}$ and target state $\ket{f}$ are indirectly coupled via some intermediate excited state, $\ket{e}$ as shown in Fig.~\ref{fig1}. 
The states $\ket{g}$ and $\ket{f}$ are here considered to be `quasi-stable' ground states, where $\ket{e}$ is a radiatively decaying excited state. The typical Hamiltonian for this physical system reads 
\begin{equation}
	\label{Hamiltonian1}
	H(t) = \frac{\hbar}{2}\begin{pmatrix}
										0 & \Omega_P(t) & 0 \\
										\Omega_P(t) & 2\delta_P(t) & \Omega_S(t) \\
										0 & \Omega_S(t) & 2\delta(t)	
									\end{pmatrix}
\end{equation}
Here $\Omega_P(t)$ and $\Omega_S(t)$ represent the Rabi frequencies of the couplings that drive transitions $\ket{g} \to \ket{e}$ and $\ket{e} \to \ket{f}$ respectively (commonly known as the `pump' and `Stokes' couplings). The term $\delta_P = (E_e - E_g) - \omega_P$ is referred to as the `single-photon' detuning for the Pump driving field with carrier frequency $\omega_P$. The $\delta(t)$ term is the `two-photon' detuning and is defined as $\delta(t) = \delta_P - \delta_S$, where $\delta_S$ is the analogous single-photon detuning for the Stokes coupling. Control of this physical system has been extensively studied in the context of STIRAP~\cite{bergmann2019roadmap,vitanov2017stimulated,shore2017picturing}, where for $\delta\simeq0$ there exists a suitable control scheme for $\Omega_P(t)$ and $\Omega_S(t)$, the so-called {\it counterintuitive} pulse sequence, such that perfect transfer from $\ket{g}$ to $\ket{f}$ is achieved whilst $\ket{e}$ is kept depopulated at all times. Here we instead consider the case of \textit{always-on} Rabi-frequencies whilst modulating the single- and two-photon detunings. The population transfer thus achieved mimics protocols in circuit-QED where the couplings between qubit and harmonic mode are not switchable~\cite{distefano2015population}. Specifically, we investigate the case where the couplings $\Omega_P$ and $\Omega_S$ both assume the constant value $\Omega_0$, whilst freedom is afforded to modulate the detunings $\delta_P(t)$ and $\delta(t)$, which embody a set of controls of simple experimental manipulation. \belfa{The remits of our investigation extend beyond the context set by the 3-level system illustrated in this Section. Indeed, the three-level model considered here can also be used to address the problem of population transfer between two remote quantum resonators both connected by non-switchable couplings to a three-level system, which can be operated locally~\cite{falci2017advances}.
Moreover, this configuration also describes a system consisting of two qubits connected by the field of a cavity and working in the single-excitation subspace. In this context, the two low-energy states of the equivalent three-level system would represent states where a single excitation is carried by one of the remote qubits, while the top-most state would imply that the cavity field is populated. This configuration is the building block of cavity-/circuit-QED architectures for controlled quantum dynamics currently being explored experimentally.}

\section{Reinforcement Learning based optimization}
\label{ReinforcementLearning}
In order to find an efficient control scheme we first employ an RL-inspired approach. Initially, we fix the total time for the system evolution to $T$ which is then divided into $N_{steps}$ time intervals, $t_i$, of equal duration. This constitutes one episode. During each of these intervals the one- and two- photon detunings have constant values, $\delta_P(t_i)$ and $\delta(t_i)$, which are all determined by an RL agent prior to each interval. Thus, for each time interval, we use the Hamiltonian in Eq.~\eqref{Hamiltonian1} with $t\to t_i$ and $\Omega_P=\Omega_S=\Omega_0$, 
to evolve the continuous-time open-system dynamics ruled by the Lindblad master equation
	\begin{equation}
		\label{Lindblad1}
		\dot{\rho} = -\frac{i}{\hbar}\comm{H(t_i)}{\rho} + \mathcal{D}(\rho),
	\end{equation}
for the duration of the time interval. Here $\rho$ is the density matrix of the system and ${\cal D}$ is the Lindblad-like operator accounting for the non-unitary part of the dynamics. More specifically, the agent provides two values, for each individual timestep, which act as the mean values of two separate Gaussian policies from which the detuning are sampled at said timestep. Learning is implemented using the policy gradient REINFORCE (with baseline) algorithm for continuous action spaces~\cite{sutton1998reinforcement}, employing a long short-term memory (LSTM) neural network~\cite{hochreiter1997long} to as a function approximator (with only the series of time steps $\{t_i\}_{i=1,...,N_{steps}}$ as external input to the network) mirroring previous work~\cite{sgroi2020reinforcement}. 

Thus the agent is tasked with learning a policy that provides the optimal detuning control scheme, where \belfa{performance} is considered with respect to perfect transfer between $\ket{g}$ and $\ket{f}$, whilst keeping $\ket{e}$ depopulated at all times. In order to meet such a request 
we couple the intermediate state to a sink $\ket{s}$, in the learning phase only, as shown in Fig.~\ref{fig1}.
			
This coupling is operationally implemented by introducing the Lindblad operator $\sqrt{\Gamma} \op{s}{e}$ into the dissipator in Eq.~(\ref{Lindblad1}) and induces a decay mechanism in the system, whereby any protocol that appreciably populates the excited state invariably leads to population loss. \belfa{This is crucial: removing the state observation at each time-step removes the ability to explicitly define a reward function that encourages the desired dynamics.} In this case we can define the delayed reward granted to the RL agent at the end of the evolution as
\begin{equation}
\label{reward}
R = \rho_{f,f}(T).
\end{equation}
\belfa{This explicitly promotes population of the final state $\ket{f}$, whilst any transient population of $\ket{e}$ during the dynamics will act to lower this final population thanks to the aforementioned leakage mechanism. In this sense, 
punishment for populating $\ket{e}$ is built-in to the mechanisms of the system via $\ket{s}$}.

{\belfa{The way this algorithm is able to work without monitoring the system at each time step can be rationalised in the following way.} As the LSTM does not monitor the state of the system at each time step, it relies only on the ability to `memorize' the actions that it has taken at each time step leading up to the final reward $R$. Thus, over several episodes, the agent is able to build an internal representation of the system dynamics 
and thus learn to act optimally with only the series of time-steps as input and the final target-state population as feedback. Consequently this type of optimization could in principle be employed as an iterative, closed-loop scheme. Such a key feature of our approach would be beneficial for optimizing control in the presence of difficult-to-simulate environmental decoherence, such as the in the situations faced by solid-state quantum hardware~\cite{werninghaus2021leakage,elisabettaRMP}}. A detailed explanation of the RL-LSTM approach is provided in Appendix~\ref{app1}, while the network configuration --  along with all the learning parameters -- are reported in Appeendix~\ref{app2}.

\begin{figure}[b!]
	\centering
	{\bf (a)}\hskip3cm{\bf (b)}
	\includegraphics[width=1\columnwidth]{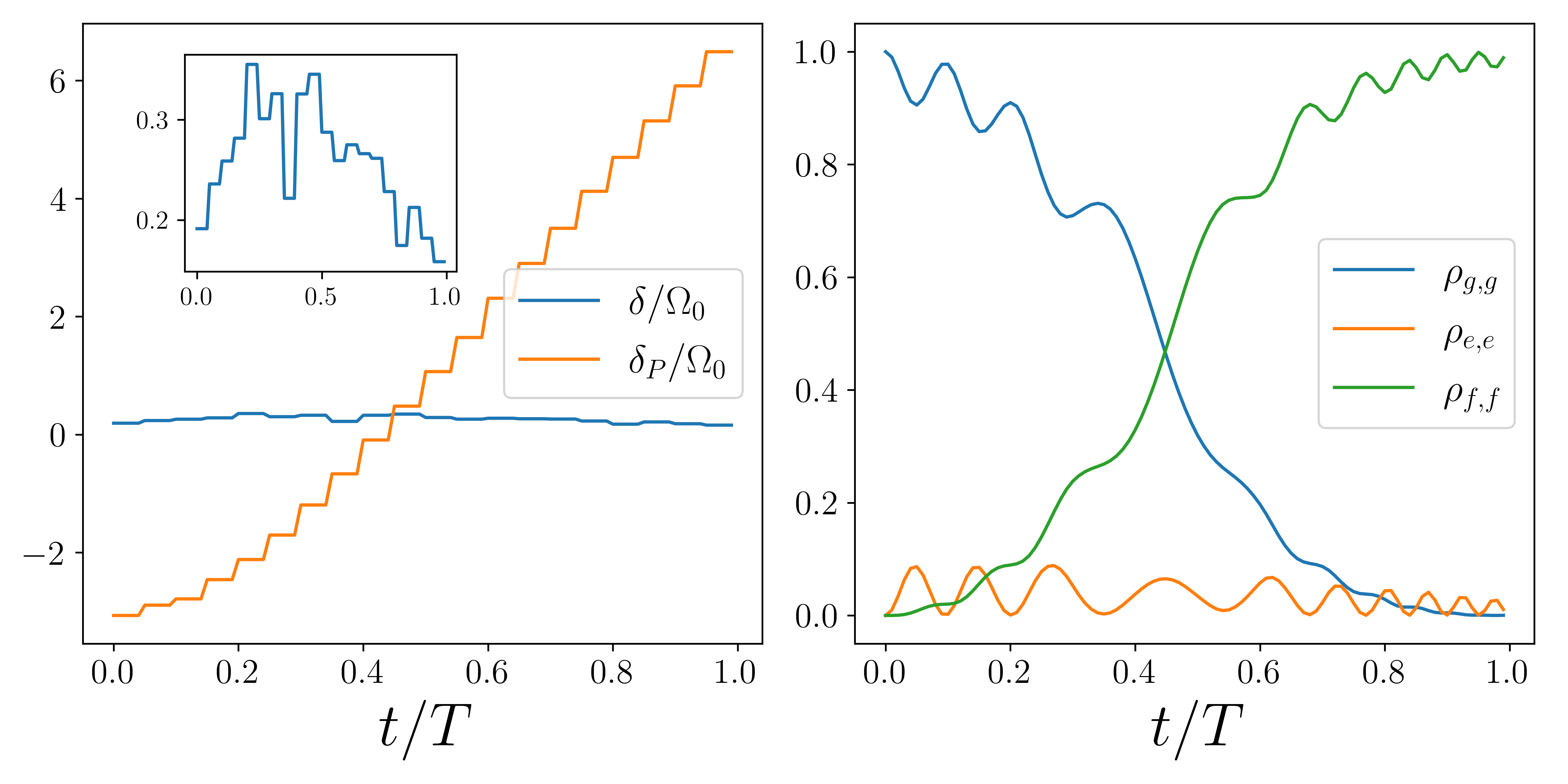}
	\caption{{\bf (a)} Detunings as a function of time for the control scheme obtained with the first RL-LSTM optimization. Notice that $\delta$ varies within a much smaller range than $\delta_P$ (cf. Inset). {\bf (b)} Population transfer achieved with the protocol shown. The target state population at the end of the protocol is $\rho_{f,f}(t) \approx 0.9993$, while the maximum excited state population is $\max_t \rho_{e,e}(t) \approx 0.0884$.}
	\label{RL_Proto_1}
\end{figure}

Using the RL based optimization outlined above, with $\Omega_0 T=20$, $N_{steps}=20$ and $({\delta}, {\delta_P})/{\Omega_0} \in [-50,50]$, the agent was able to obtain a target state population at the end of the protocol of $\rho_{f,f}(T) \approx 0.9993$, with a maximum excited state population over the entire time interval of $\max_{t} \rho_{e,e} (t) \approx 0.0884$. The learned protocol and the induced population dynamics can be found in Fig.~\ref{RL_Proto_1}.
Despite the evidently desirable features of the results thus achieved, it is worth remarking that the learning process is in general stochastic and different runs of the optimization can produce different shapes for the detuning functions. However, successfully optimized detuning functions all shared common traits, which can be summarized by the following list of characteristic features
\begin{itemize}
	\item[{\bf C1:}] We have $\abs{\delta(t)} \ll \abs{\delta_P(t)}$ for most of the evolution.
	
	\item[{\bf C2:}] Detuning $\delta_P(t)$ always exhibits comparatively large initial and final values.
	
	\item[{\bf C3:}] $\delta_P(t)$ always seems to exhibit specific parity features about $T/2$. Such a feature is more sporadically shared by $\delta(t)$. 
\end{itemize}
In particular, feature {\bf C2} is to be expected if one wants to avoid populating the excited state at the beginning and at the end of the transfer, and agrees with previous findings reported in literature~\cite{distefano2015population}. Furthermore, feature {\bf C1} can be justified by inspecting how the presence of non-vanishing detunings affects the efficiency of both standard STIRAP and Raman protocols: while even small non-null values of  $\abs{\delta}$ 
are detrimental for the performance of the transfer,  much larger values of $\delta_P$ can be tolerated~\cite{falci2013design,vitanov2017stimulated,shore2017picturing}. 

\begin{figure}[t!]
	\centering
{\bf (a)}\hskip3cm{\bf (b)}
\includegraphics[width=1\columnwidth]{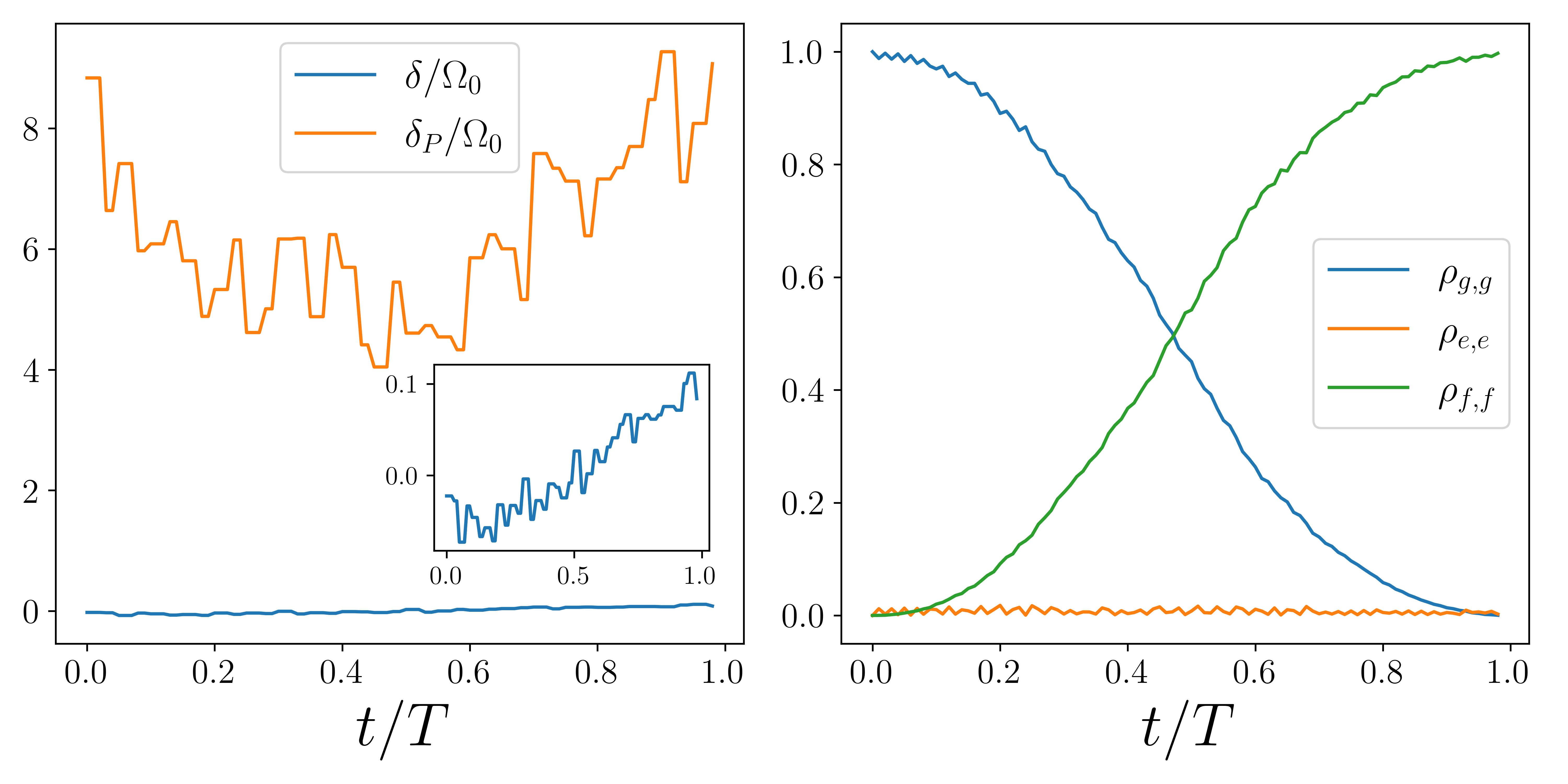}
	\caption{{\bf (a)} Detunings as functions of time for the control scheme obtained with the second RL-LSTM optimization. {\bf (b)} Population transfer achieved with the protocol shown. The target state population at the end of the protocol is $\rho_{f,f}(t) \approx 0.9999$, while the maximum excited state population is $\max_t \rho_{e,e}(t) \approx 0.0179$.}
	\label{RL_Proto_2}
	\label{RL_Dyna_2}
\end{figure}

We have performed an optimization process based on the use of a restricted range for the values of $\abs{\delta}$, thus limiting the action-space of the RL agent and guaranteeing the validity of {\bf C1}. In particular, we considered ${\delta}/{\Omega_0} \in [-0.2, +0.2]$ and ${\delta_P}/{\Omega_0} \in [-14, +14]$. We have also taken a longer evolution time $\Omega_0 T = 40$, with proportionally more steps whereby the agent can act ($N_{steps} = 40$). The resulting protocol can be seen in Fig.~\ref{RL_Proto_2} {\bf (a)}. The RL agent obtained a maximum end-protocol target state population of $\rho_{f,f}(T) \approx 1-10^{-4}$, with a maximum transient excited state population of $\max_{t} \rho_{e,e} \approx 0.0179$. The corresponding dynamics is shown in Fig.~\ref{RL_Dyna_2} {\bf (b)}.

{Remarkably, differently from what one would naively expect, this protocol is not akin to a Raman-like or a 
STIRAP-like one. 
First, two-photon Raman protocols require large single-photon detunings while, in our case, $\delta_P$ can even vanish, thus making the dynamics comparatively faster. Second, the protocol that we have found are non-adiabatic, thus making them markedly different from  adiabatic population transfers, such as STIRAP. Our LSTM RL approach thus delivers genuinely new protocols that combine features of robustness akin to STIRAP but without requiring the demanding switching of coupling fields}

\section{Polynomial Coefficient Optimization}
\label{CoeffOptimization}
Instead of dividing the time of the evolution in a certain number of steps and optimizing the values of the detunings at each step, an alternative approach for the optimization consists on the expansion of $\delta(t)$ and $\delta_P(t)$ over a specific functional basis. The effectiveness of this approach depends on the choice of such basis, making it less general than the technique used in the previous section or other sophisticated optimal control techniques such as CRAB~\cite{caneva2011chooped}. However, should a suitable basis be found, the suggested approach translates the problem of finding the best protocol into a simpler numerical optimization over the coefficient of the expansion while also providing us with a simple analytical expression for the control terms.

\begin{figure}[b!]
	\centering
	{\bf (a)}\hskip3cm{\bf (b)}
	\includegraphics[width=\columnwidth]{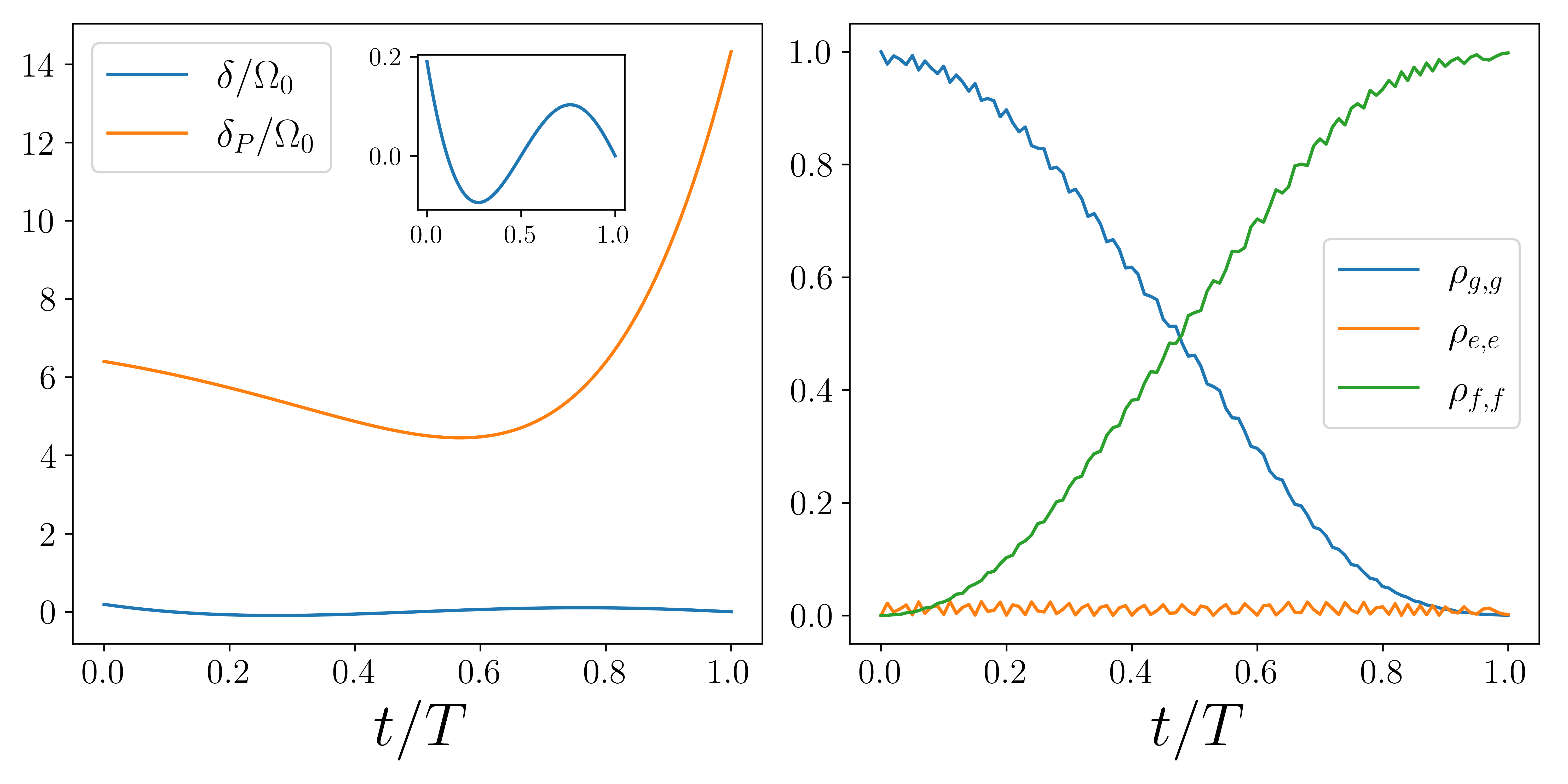}
	\caption{\small{{\bf (a)} Example of optimized protocol obtained using a polynomial basis-expansion for the optimization of the detunings. The inset shows the behavior of the respective two-photon detuning in a smaller vertical range. {\bf (b)} Corresponding population transfer. The target state population at the end of the protocol is $\rho_{g2,g2}(T)\approx0.9980$, while the maximum excited state population is $max_t\rho_{e,e}(t)\approx0.0250$}.}\label{poly_ex1}
\end{figure}

\begin{figure}
	\centering
	{\bf (a)}\hskip3.5cm{\bf (b)}
	\includegraphics[width=\columnwidth]{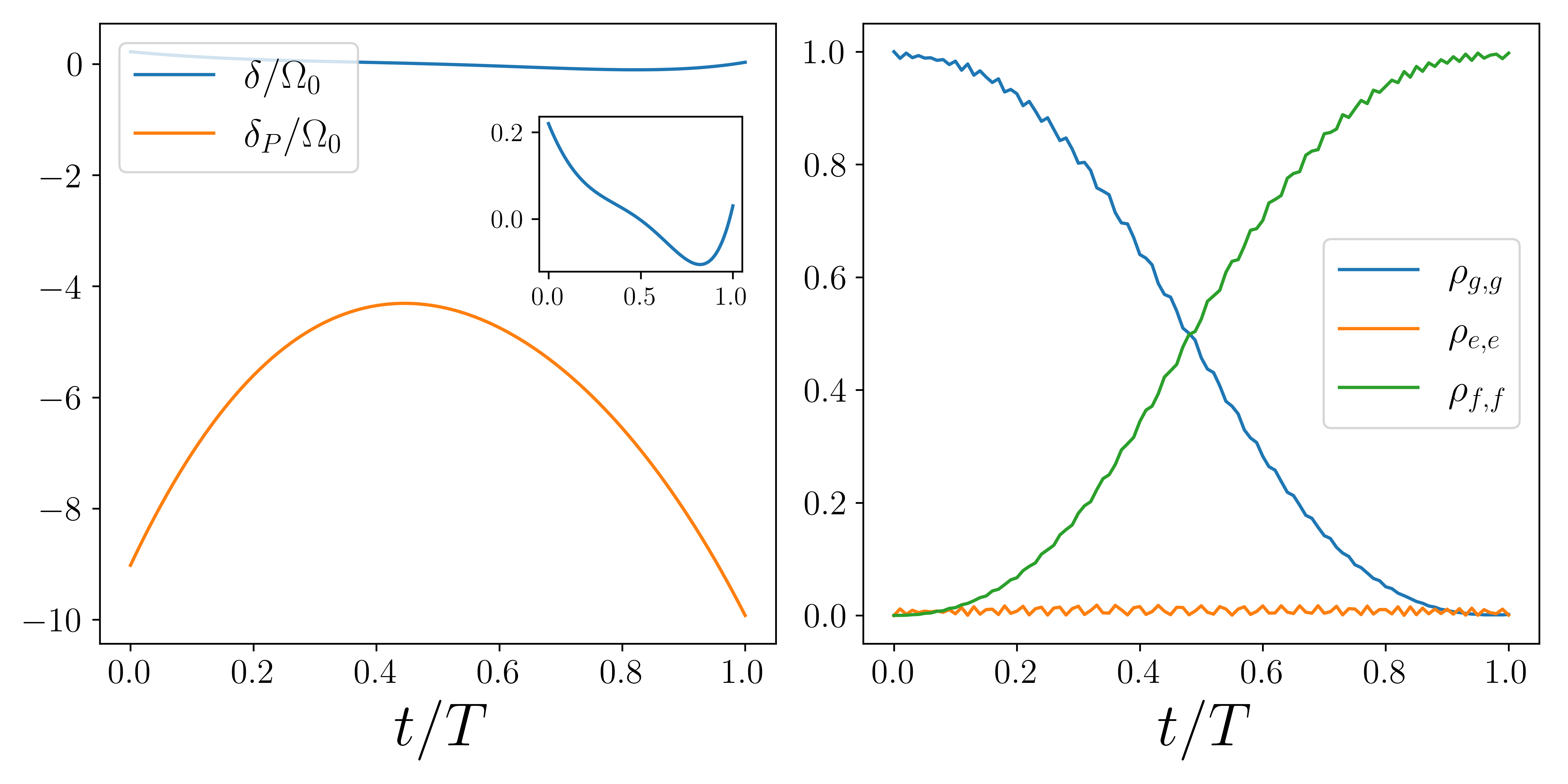}
	\caption{{\bf (a)} Example of polynomial-basis optimization protocol for the detunings. Inset: Temporal behavior of the two-photon detuning in a magnified vertical scale. {\bf (b)} Resulting performance of population transfer. The target-state population at the end of the protocol is $\rho_{g2,g2}(T)\approx0.9991$, while the maximum excited-state population is $max_t\rho_{e,e}(t)\approx0.0180$.}\label{poly_ex2}
\end{figure}

We found that writing $\delta(t)$ and $\delta_P(t)$ as $5^{th}$ order polynomial functions and using a Powell method search~\cite{2020SciPy-NMethNO} over the coefficients of the polynomial expansion is enough to achieve an effective population transfer. In Figs.~\ref{poly_ex1} and \ref{poly_ex2} we show the best protocols obtained after $10$ different runs of the optimization for $\Omega_0 T=40$. It can be seen that, while still effective, they are different from the protocol found via the RL-based optimization (although conditions {\bf C1} and {\bf C2} found by the RL agent can still be observed). This again suggests that various quasi-optimal protocols can be identified as candidates for an efficient population transfer.However, the effectiveness of such optimization technique depends on the choice of the basis for the specific problem. Performing a simple numerical optimization to solve the same problem assigned to the RL agent (finding the values of piecewise constant functions) gives us far worse solutions compared to those obtained using the the RL-based approach ~\cite{sgroi2020reinforcement}.
Therefore, not only the RL-based approach can be successfully applied to a wider class of problems with a simpler pre-optimization analysis but it also provides a better exploratory tool when only sub-optimal solutions are achieved, as these solutions are not biased by the choice of a specific basis of functions.

\section{Optimal Protocols}
\label{OptimalProtocols}
Based on the success of both the RL-based optimization and the optimization with a polynomial basis, we combined the two approaches, performing a straightforward numerical optimization starting from the results of the RL-based technique. To this end, observing the features in Fig.~\ref{RL_Proto_2}, we propose an \textit{ansatz} for $\delta_P(t)$ as
\begin{equation}
	\frac{\delta_P(t)}{\Omega_0} = C_1 - C_2 e^{k \left(\frac{t}{T} - 0.5\right)^2}.
	\label{dp2}
\end{equation}
Similarly, we suggest the linear {ansatz} for $\delta(t)$,
\begin{equation}
	\frac{\delta(t)}{\Omega_0} =  m\left(\frac{t}{T} - 0.5\right).
	\label{linear}
\end{equation}
The choice of Eqs.~\eqref{dp2} and \eqref{linear} ensure that the symmetry or anti-symmetry point of the proposed functions occur at $t={T}/{2}$ of the  evolution.

\begin{figure}[ht]
	\centering
	{\bf (a)}\hskip3.5cm{\bf (b)}
	\includegraphics[width=1\columnwidth]{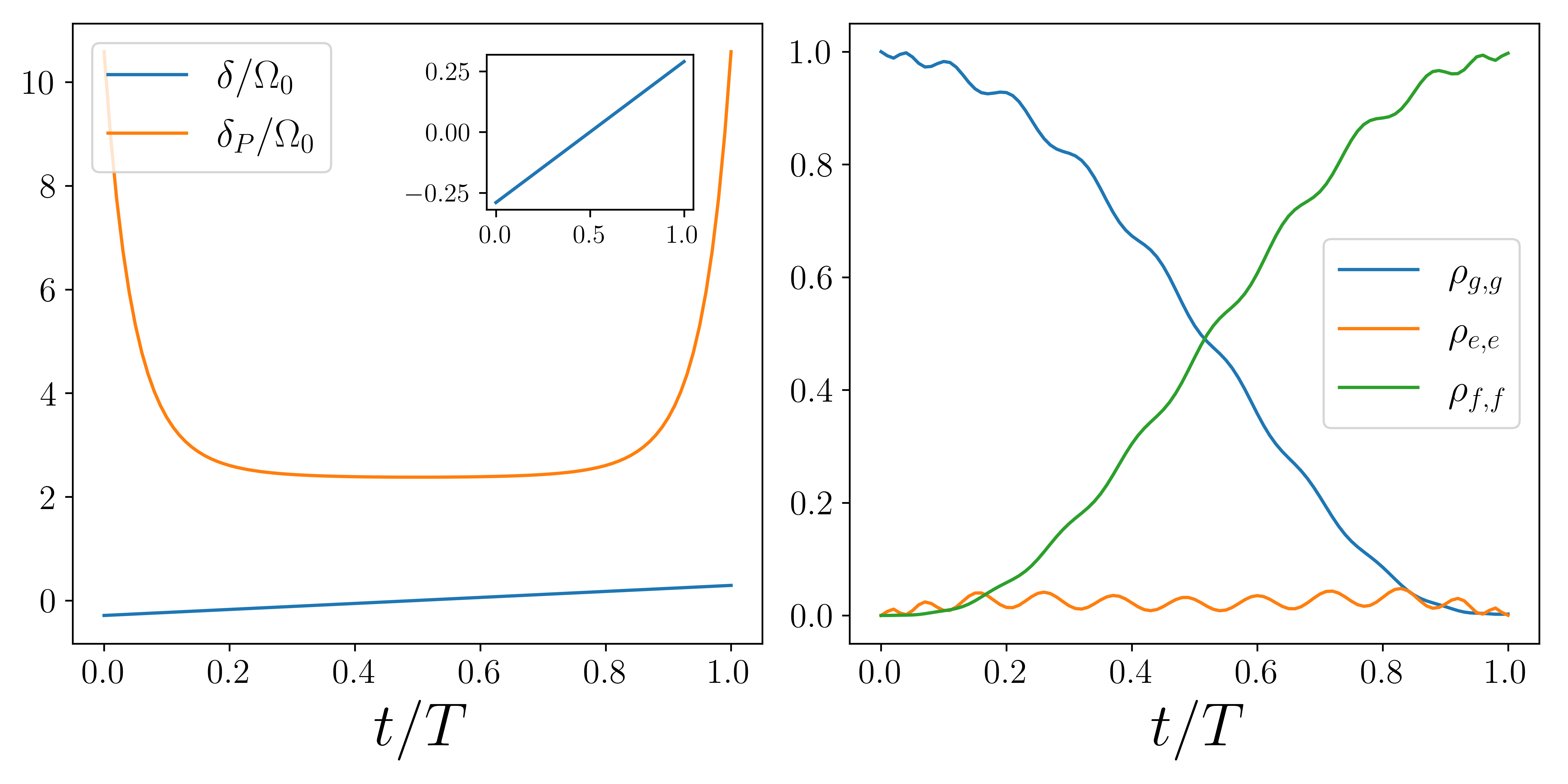}\\
		{\bf (c)}\hskip3.5cm{\bf (d)}
			\includegraphics[width=1\columnwidth]{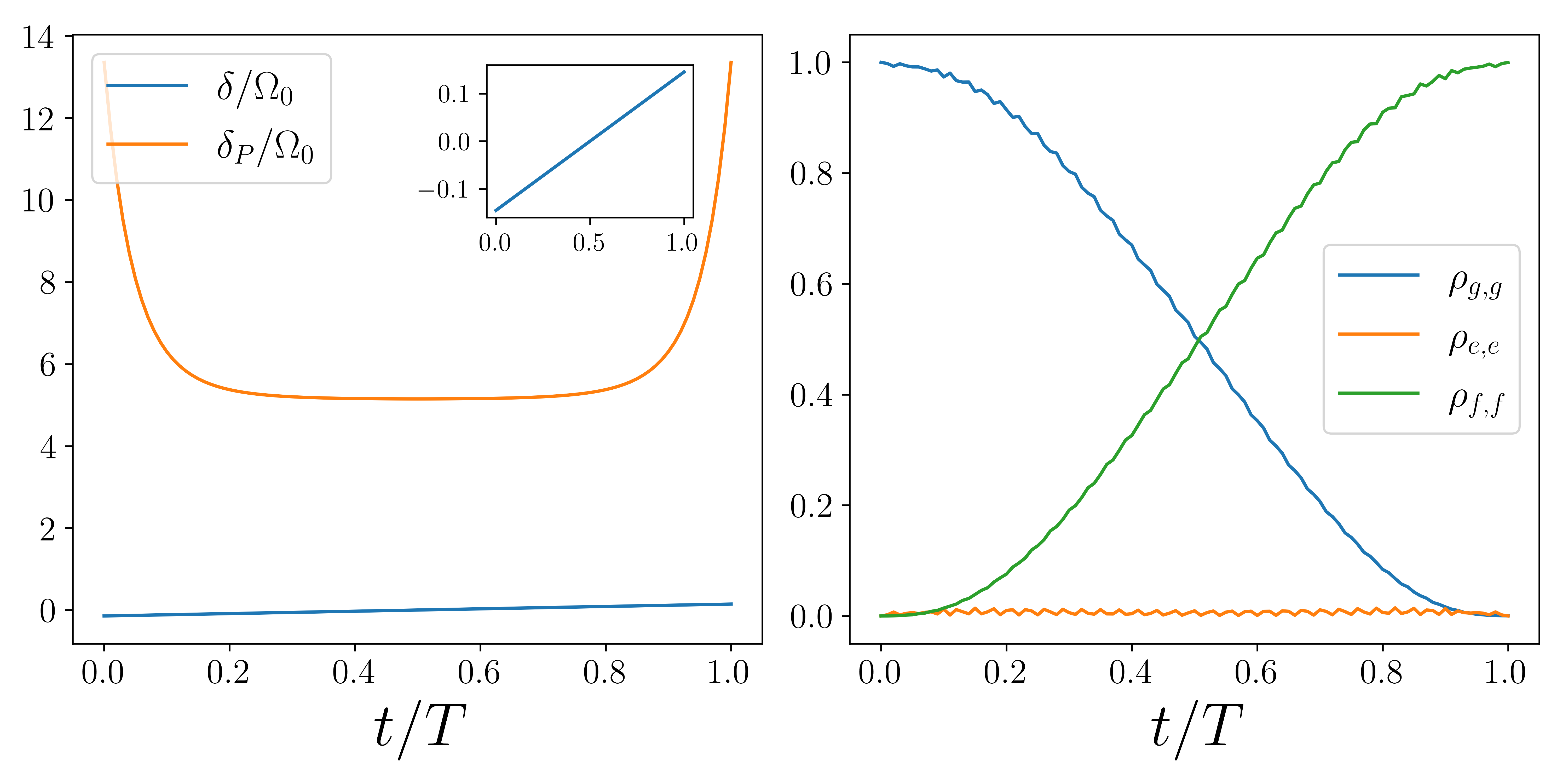}
	\caption{{\bf (a)} Optimized detunings for protocol 1 [cf. Eqs.~\eqref{dp2} and  \eqref{linear}] for $\Omega_0T=20$. The optimal parameters are $C_1\approx2.34$, $C_2\approx-0.038$, $k\approx21.52$, $m\approx0.58$. 
	The feature $|\delta(t)|<<\delta_P(t)$ is satisfied at all times, including in the region where $\delta_p(t)$ has an approximately constant trend. {\bf (b)} Optimized population dynamics corresponding to the protocol in panel {\bf (a)}. The target state population at the end of the protocol is $\rho_{f,f}(T)\approx0.9968$, while the maximum excited state population is $\max_t\rho_{e,e}(t)\approx0.0476$. {\bf (c)} Same as in panel {\bf (a)} but for $\Omega_0T=40$. We have $(C_1,C_2,k,m)=(5.11,0.038,21.51,0.29)$ 
	{\bf (d)} Optimized population dynamics corresponding to the protocol in panel {\bf (c)}. The target-state population is $\rho_{f,f}(T)\approx0.9994$, while the maximum excited-state population is $\max_t\rho_{e,e}(t)\approx0.0143$.}

	\label{RL_Proto_3&4}
\end{figure}

This optimization was carried out using a {Powell method} search~\cite{2020SciPy-NMethNO} over the space of parameters $(C_1, C_2, k,m)$ for the maximization of $R$. The benefit here is two-fold: on one hand, it allows us to find an analytical expression for the protocol, thus contributing to the interpretation of the results that we achieve; on the other hand, it smooths the protocol found by the RL agent, presenting us with a continuous control scheme, which is experimentally more tractable. In achieving these two goals the analytic, smooth control pattern maintained a comparable final state target population to the RL learned scheme, while further reducing the transient population of the excited state. Specifically, in Fig.~\ref{RL_Proto_3&4} we present results for $\Omega_0 T = 20$ and $\Omega_0 T = 40$ showing that, for the second case, $\rho_{f,f}(T) \approx 0.9994$ and $\max_{t} \rho_{e,e}(t) \approx 0.0143$ can be achieved with the simple ansatz that we have proposed. 

\begin{figure}[ht]
    \centering
    	{\bf (a)}\hskip3.5cm{\bf (b)}
	\includegraphics[width=1\columnwidth]{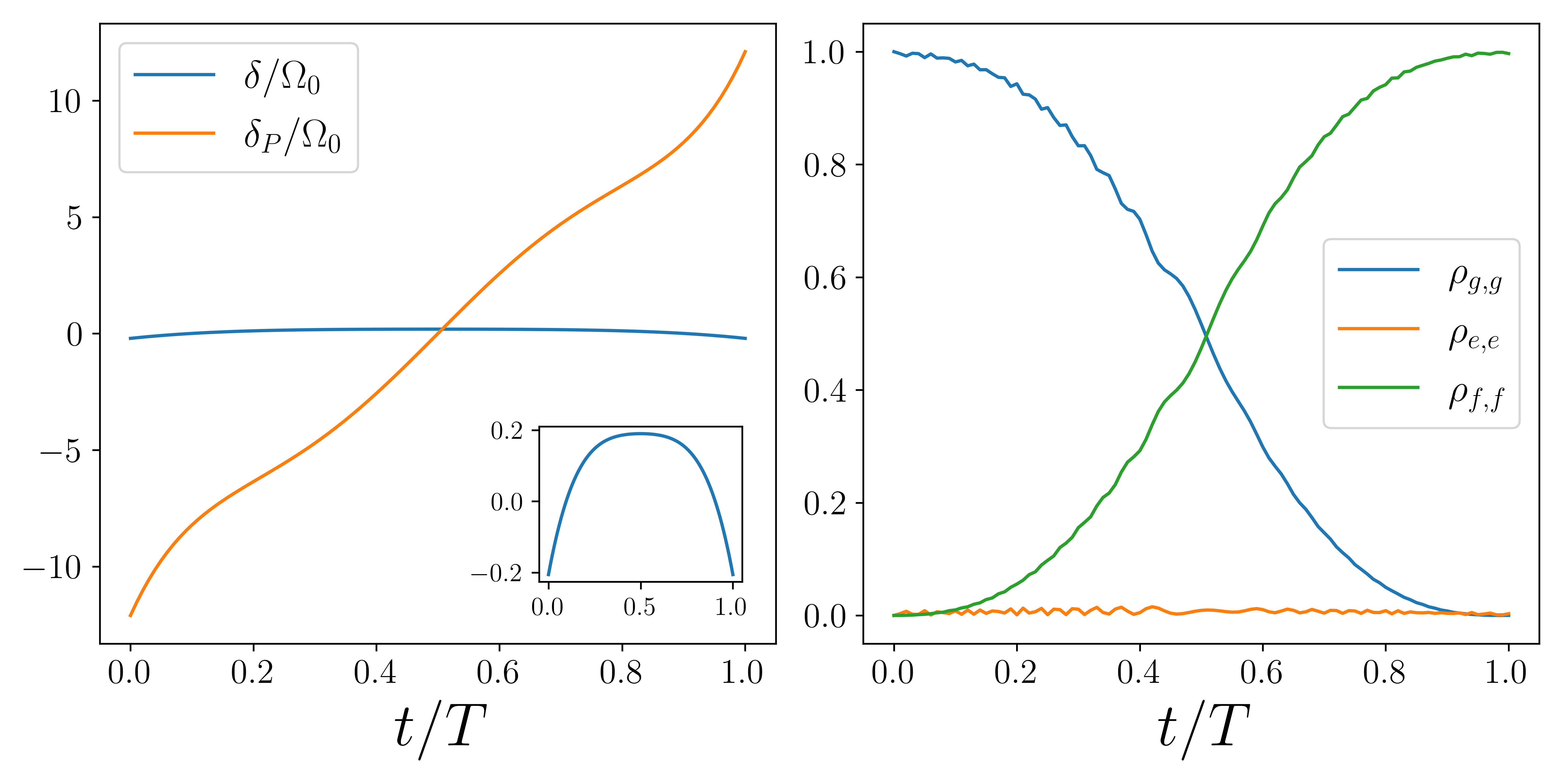}
	\caption{{\bf (a)} Optimized functional form of the detunings for protocol 2 for $\Omega_0T=40$ and the choice of an even $4^{th}$-order polynomial function for $\delta(t/T-0.5)$ and an odd $5^{th}$-order one for $\delta_P(t/T-0.5)$. The resulting protocol involves the the functions $\delta(t)/\Omega_0\approx0.19-0.37(t/T-0.5)^2-4.85(t/T-0.5)^4$ and $\delta_P(t)/\Omega_0\approx26(t/T-0.5)-87(t/T-0.5)^3+312(t/T-0.5)^5$. {\bf (b)} Resulting optimized population dynamics. The target-state population is $\rho_{f,f}(T)\approx0.9969$, while the maximum excited-state population is $\max_t\rho_{e,e}(t)\approx0.01555$.}
	\label{RL_Proto_5}
\end{figure}

The insight provided by the RL-based optimization approach suggests the existence of different valid protocols of optimization. In this regard, an interesting question to pose addresses the role of the parity exhibited by the detuning functions with respect to ${t} = T/2$. That is, we wonder whether optimal functional behaviors akin to those exhibited in Fig.~\ref{RL_Proto_1} can be identified. To ascertain it, we propose the use of an odd $5^{th}$ order polynomial function for $\delta_P\left(\frac{t}{T} - 0.5\right)$ and an even $4^{th}$ order polynomial for $\delta\left(\frac{t}{T} - 0.5\right)$ and performed a similar optimization, finding that the corresponding optimized protocol is still effective [cf. Fig.~\ref{RL_Proto_5}]. The resulting final target-state population is $\rho_{f,f}\approx 0.9969$, while the maximum excited-state population is $\max_t \rho_{e,e}(t) \approx 0.01555$. For brevity, we label the protocol of Fig.~\ref{RL_Proto_3&4} {\bf (c)}-{\bf (d)} as \textit{protocol 1}, while that of Fig.~\ref{RL_Proto_5} will be referred to as \textit{protocol 2}. \belfa{We point out that the performances of \textit{protocol 1} and \textit{protocol 2} mentioned here are extremely similar to the RL protocol of Fig.~\ref{RL_Proto_2}. Optimality can thus be understood in terms of the evident simplicity of the control functions needed to achieve such performance}

\begin{figure}[b!]
    \centering
    	{\bf (a)}\hskip3.5cm{\bf (b)}
    \includegraphics[width=\columnwidth]{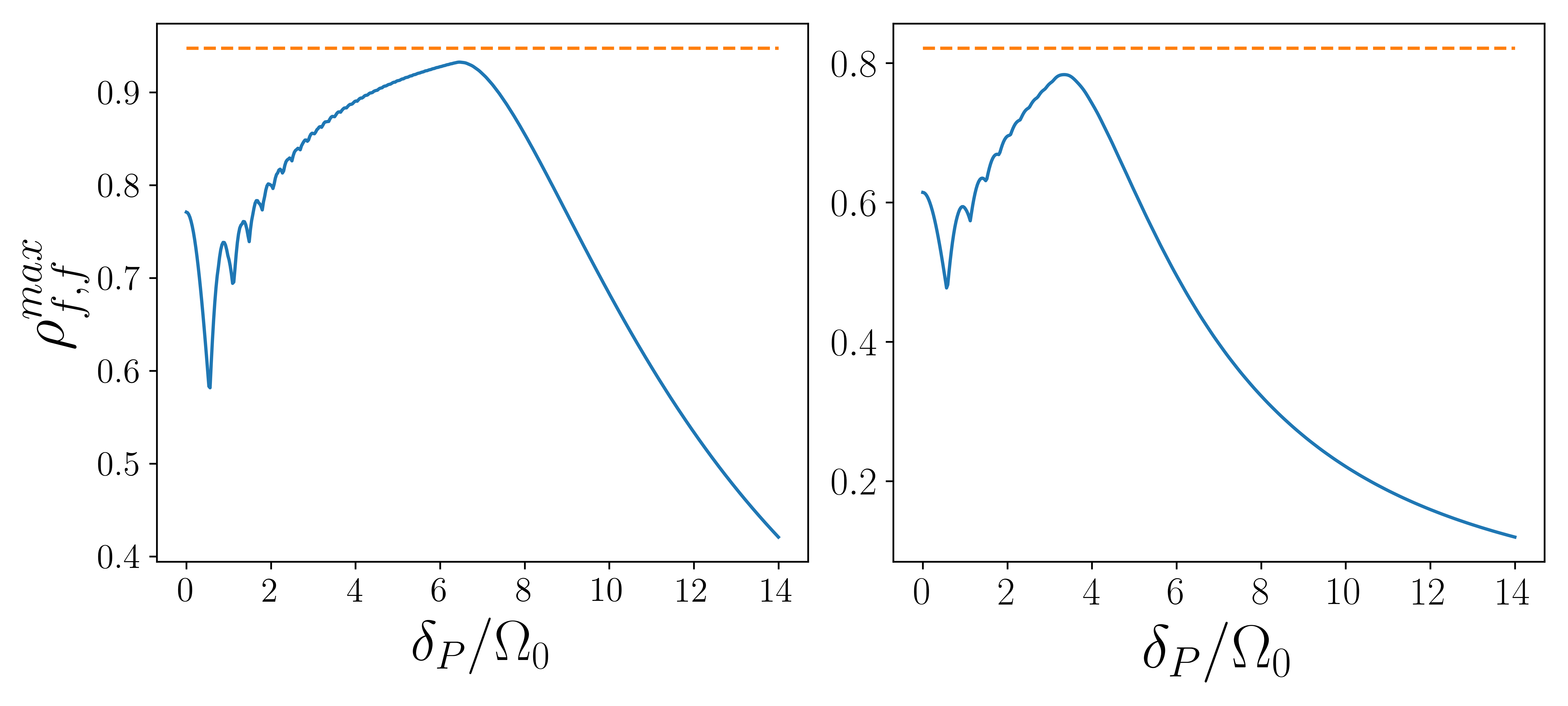}
    \caption{Maximum target-state population reached during the transfer performed using a protocol with $\delta=0$ and a constant value of $\delta_P$ for the system in Fig.~\ref{RL_Proto_1}. The dashed horizontal line indicates the population of the target state achieved using {\it protocol 1}. In panel {\bf (a)} [panel {\bf (b)}] we have used $\Omega_0 T= 40$ [$\Omega_0 T= 20$].}
    \label{adiabatic_elim_comparison}
\end{figure}

\begin{figure}[ht]
    \centering
    {\bf (a)}\hskip3.5cm{\bf (b)}
    \includegraphics[width=\columnwidth]{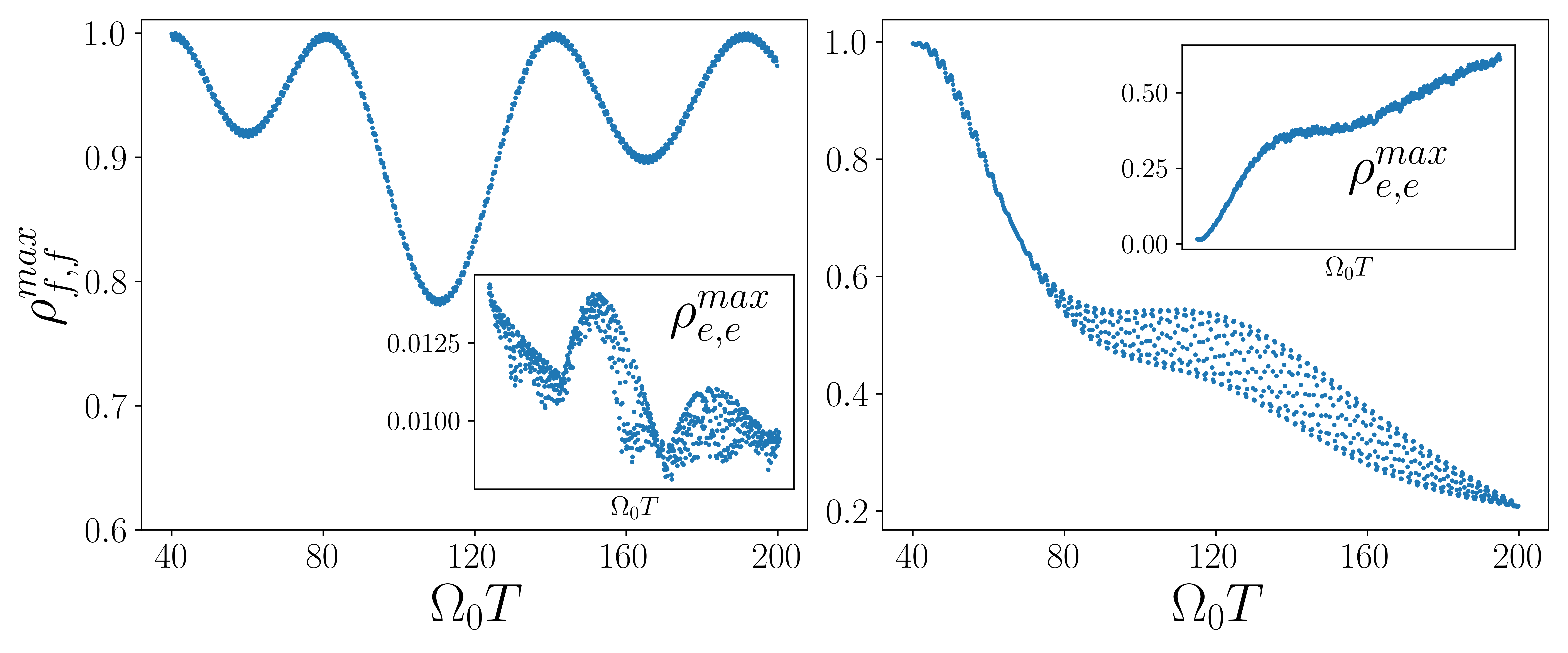}
    \caption{Target-state population and maximum excited-state population (insets) achieved with {\it protocol 1} [panel {\bf (a)}] and {\it 2} [panel {\bf (b)}] as a function of $\Omega_0T$. Both protocols are optimized for $\Omega_0T=40$. For Raman-like protocols, one would expect oscillations with a period $\Omega_0T\sim40$, which our results do not exhibit, thus marking the difference between these approaches. }
    \label{adiabaticity}
\end{figure}

The behaviours showcased in our results allow us to corroborate quantitatively the differences between our protocols and Raman-like ones. The first clear difference is the absence of Raman oscillations [cf. Fig.~\ref{adiabaticity}] from the dynamics of the populations resulting from our protocols. A second difference between the two approaches stems from the fact that in {\it protocol 1} $\delta_P$ is constant most of the time and we get $\delta(t)\ll\delta_P(t)$. One can then ask how this compares to a Raman scheme 
with $\delta=0$ and a constant $\delta_P\gg\Omega_0$. When no constraint is imposed over the total time of the evolution, one would expect that increasing $\delta_P$ will progressively improve the transfer. However, our approach assumes a fixed value of $\Omega_0 T$. This implies that a very large value of $\delta_P$ could prevent the completion of the corresponding very slow population transfer. Both of these effects are relevant for the optimal choice of $\delta_P$. In Fig.~\ref{adiabatic_elim_comparison} we show that {\it protocol 1} achieves a more efficient population transfer relative to the case of a completely constant Hamiltonian.
Moreover, in line with previous considerations, we also remark that the protocols are not adiabatic. If we increase the total time of the evolution while still using {\it protocol 1} and {\it 2} (without performing a new optimization for each value of $\Omega_0 T$), the performance does not increase monotonically, as it would happen in a Raman protocol (Figure \ref{adiabaticity}).

\section{Resilience to Decoherence}
\subsection{Spontaneous Decay}
\label{Decay}
\begin{figure}[b!]
    \centering
    {\bf (a)}\hskip3.5cm{\bf (b)}
    \includegraphics[width=\columnwidth]{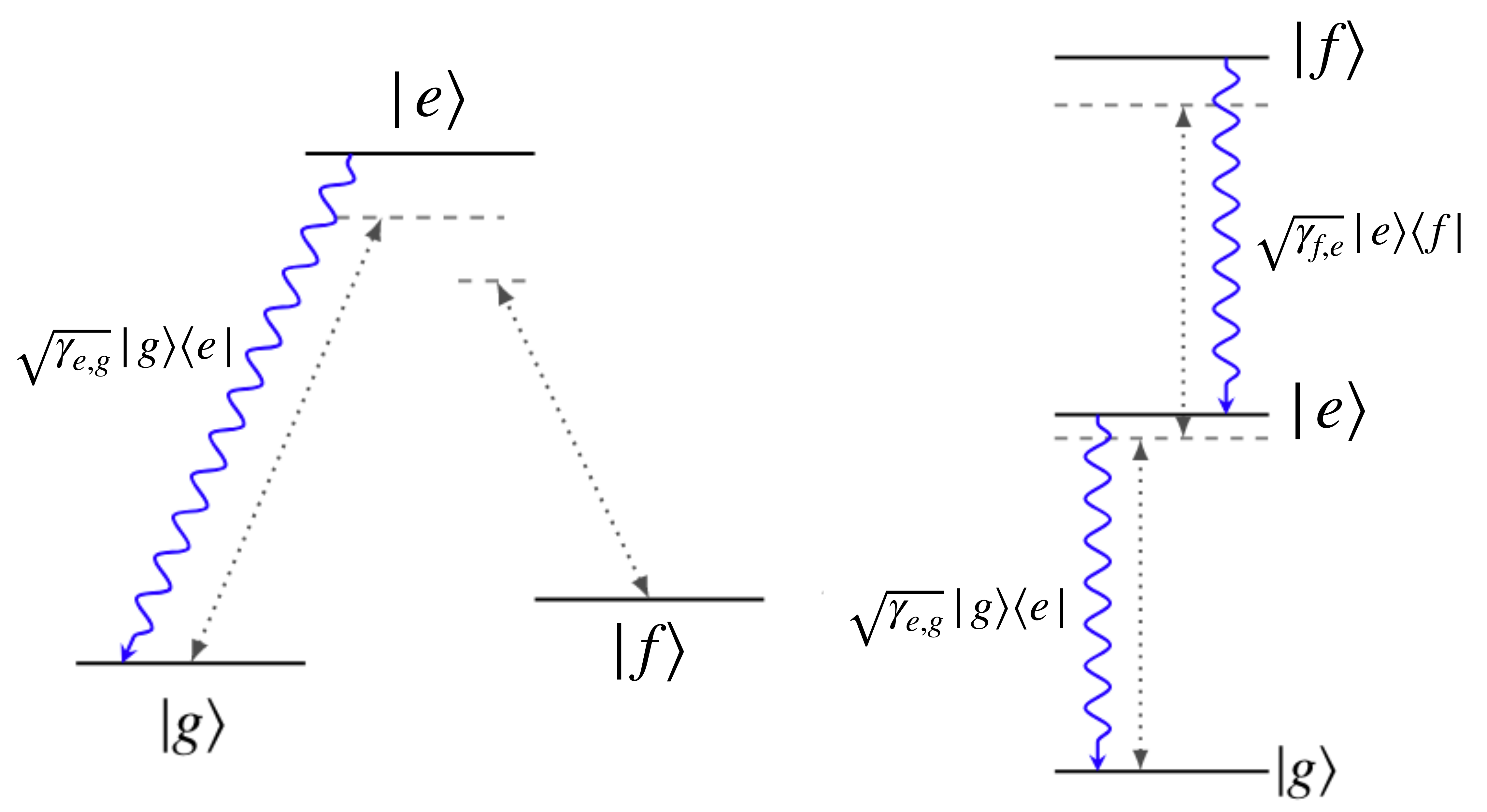}
	\caption{Schematics of three-level systems affected by spontaneous decay. We consider: {\bf (a)} a Lambda system with intermediate-level decay only; {\bf (b)} a Ladder scenario where both the intermediate state and the target state are subjected to decay.}
	\label{Decay_fig}
\end{figure}

Here we consider how the protocols that we have found  perform when a multi-level system is subjected to spontaneous decay from some of its energy levels. We investigate two cases:
\begin{enumerate}
\item[ {\bf A}:] The decay from the intermediate excited state $\rho_{e,e}$ to the first ground state $\rho_{g,g}$ with a decay rate $\gamma_{e,g}$, implemented using the Lindblad operator $\sqrt{\gamma_{e,g}}\op{g}{e}$. 
\item[ {\bf B}:] The case of an additional decay channel, from $\rho_{f,f}$ to $\rho_{e,e}$, with rate $\gamma_{f,e}$, implemented by $\sqrt{\gamma_{f,e}}\op{e}{f}$. 
\end{enumerate}
Scenario {\bf A} is what one would expect to be relevant for the Lambda system that we have discussed thus far, particularly when fluxonium-based embodiments of the multi-level system are considered~\cite{fluxonium}, for which an incoherent mechanism driving decay of population from $\ket{f}$ to $\ket{g}$ can be safely neglected. On the other hand, scenario {\bf B} is motivated by the fact that the Hamiltonian in Eq.~\eqref{Hamiltonian1} encapsulates the so-called Ladder energy level structure, where $\ket{f}$ becomes a higher excited state than $\ket{e}$ and is thus  susceptible to spontaneous decay. The Lambda and Ladder scenarios are operationally equivalent as far as the control protocols are concerned. A diagrammatic outline of these two scenarios is shown in Fig.~\ref{Decay_fig}.

\begin{figure}[t!]
\centering
{\bf (a)}\hskip3.5cm{\bf (b)}
	\includegraphics[width=\columnwidth]{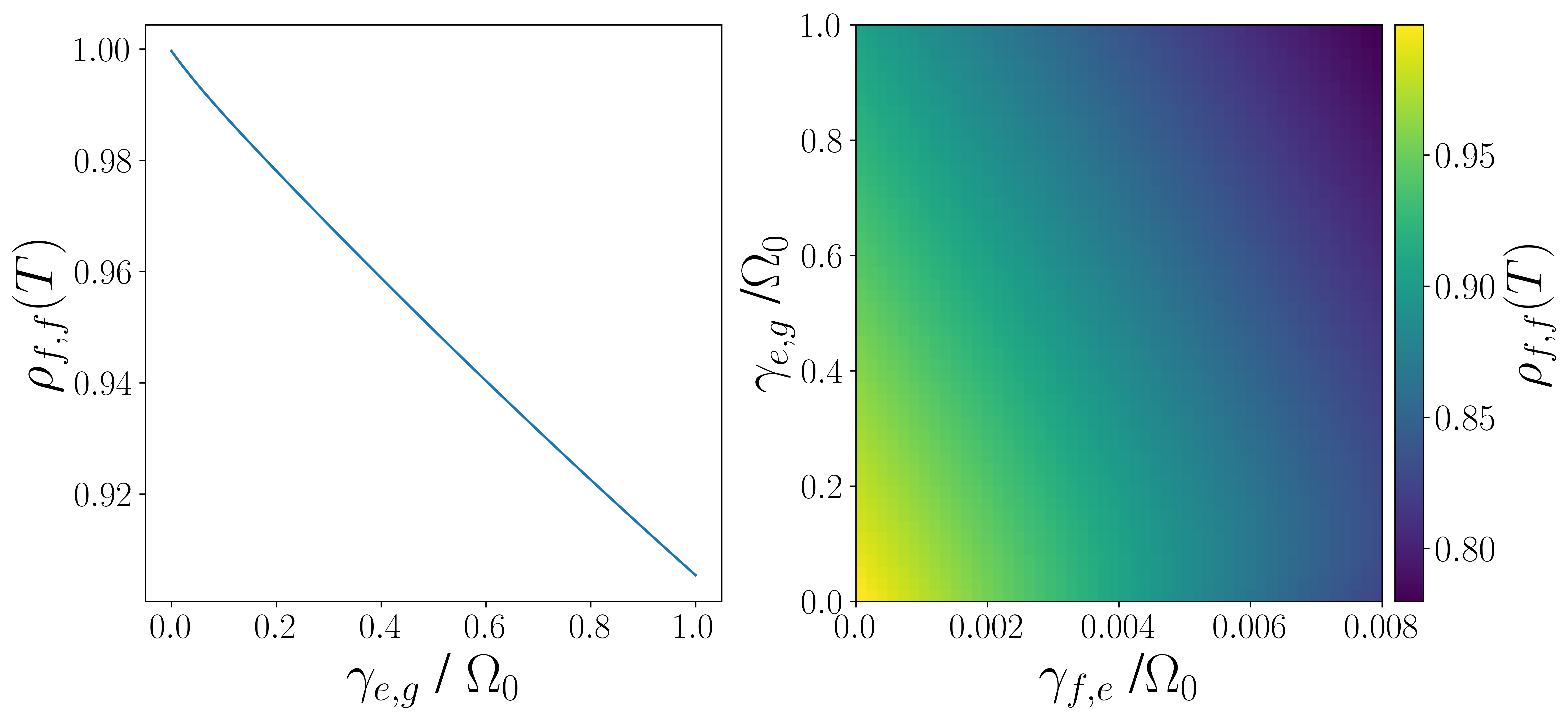}
	{\bf (c)}\hskip3.5cm{\bf (d)}
		\includegraphics[width=\columnwidth]{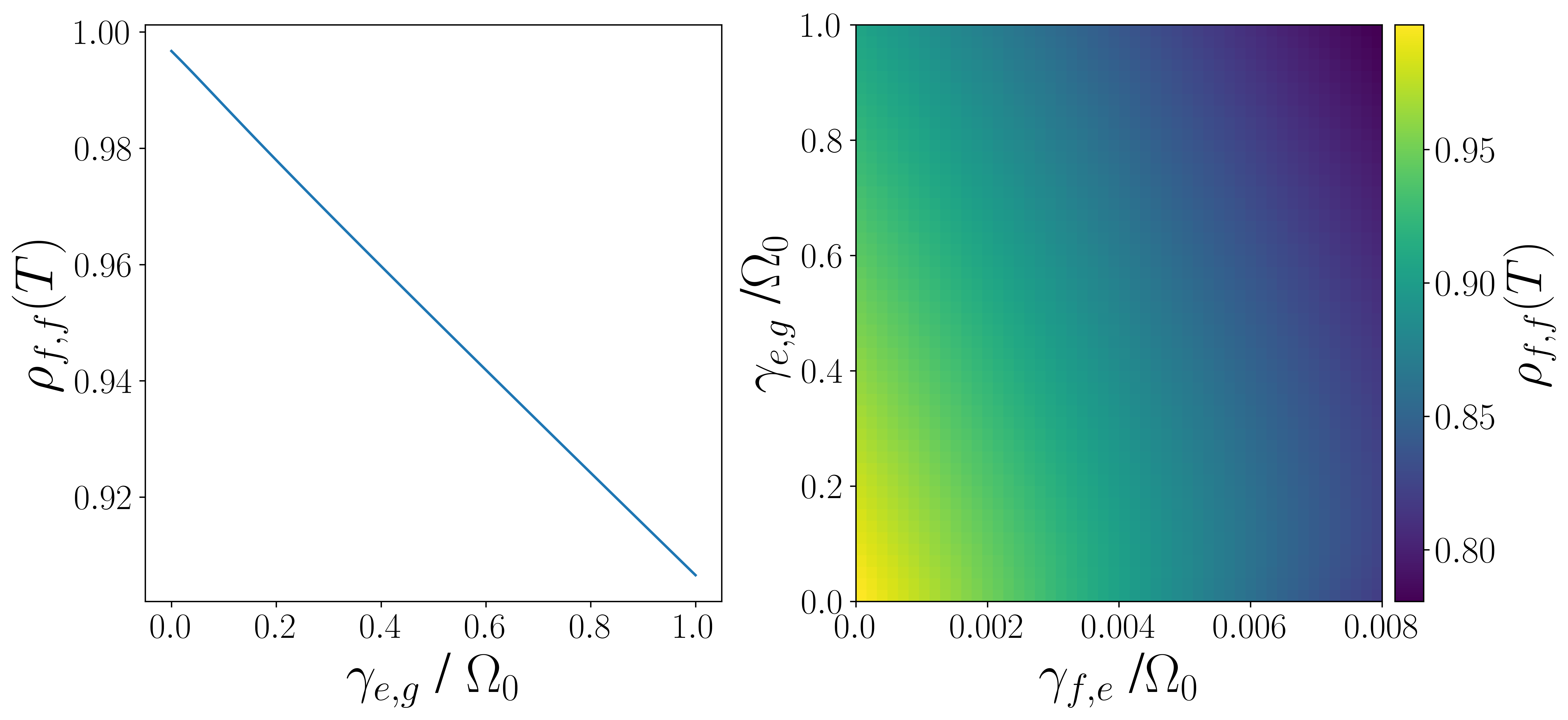}
\caption{{\bf (a)} Sensitivity of the performance for {\it protocol 1} in the presence of the single-level decay mechanism. {\bf (b)} Same analysis performed against the two decaying levels of the Ladder scheme. Note how the ranges used for the two decay rates vastly differ, this is a consequence of the protocols being, by construction, considerably more robust to decay from the intermediate excited state. In panel {\bf (c)} [panel {\bf (d)}] we show the sensitivity of the performance of {\it protocol 2} under the effects of the single-channel [double-channel] decay mechanism.}
\label{Sensitivities1}
\end{figure}

We consider the sensitivity of the protocols, with respect to final target state population $\rho_{f,f}(T)$, for a range of decay rates in both cases. Fig.~\ref{Sensitivities1} shows how the performance of protocols 1 and 2 respectively depend on the strength of the decay rates in each of the Lambda and Ladder cases, where performance is gauged simply by the final target state population.

It can be appreciated that both protocols carry a strikingly similar dependence on the decay rates and exhibit relative robustness against decay from the intermediate state. This is to be expected the RL process included a mechanism to punish population of such level. On the other hand, both protocols exhibit great sensitivity to decay from the target state. Thus, in a ladder system, a decaying target state embodies the main limiting factor.

\subsection{Dephasing}
\label{Dephasing}
\begin{figure}[t!]
    \centering
    \includegraphics[width=0.8\columnwidth]{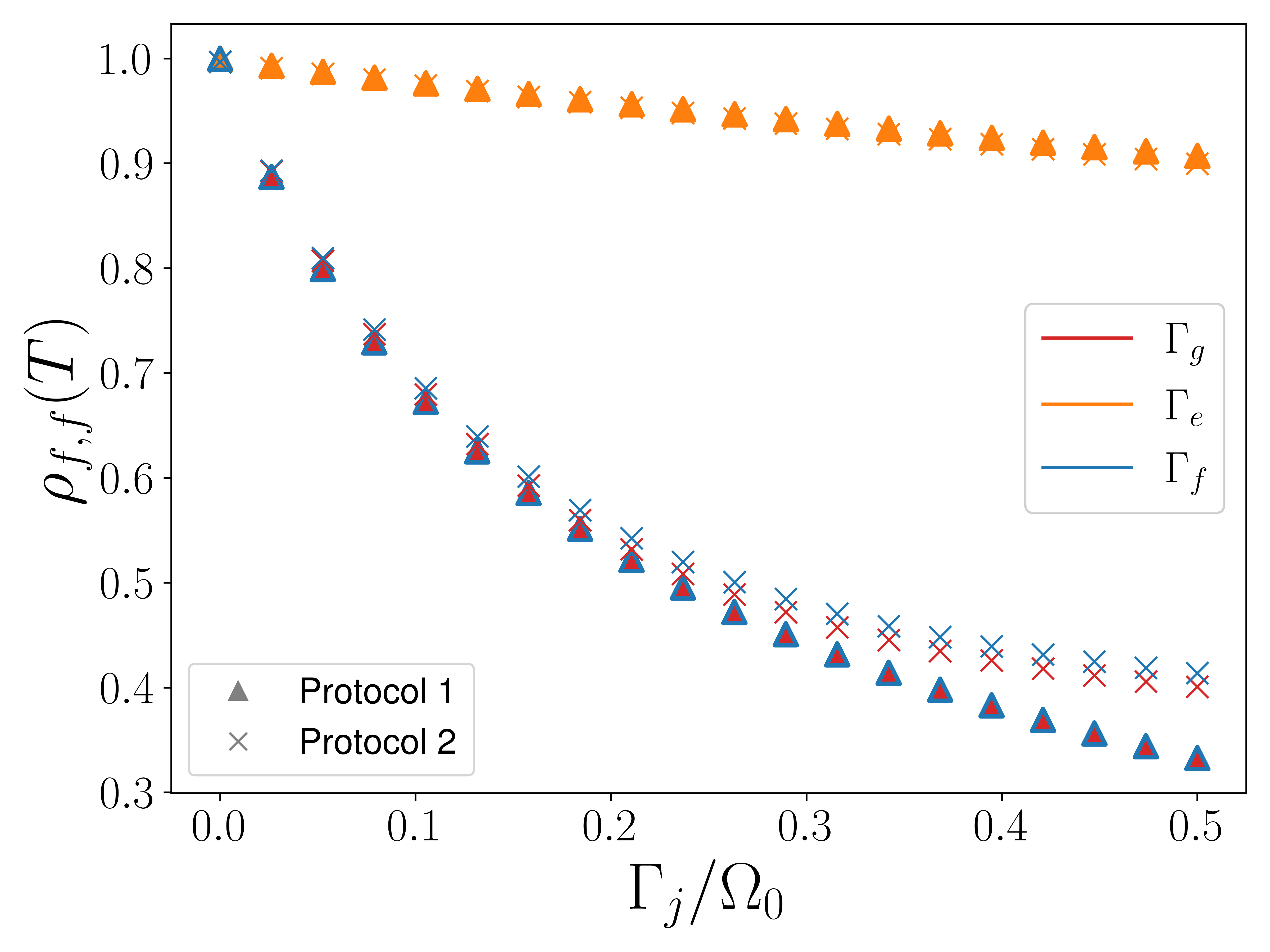}
    \caption{Final population achieved by the optimal protocols while independently varying the dephasing strengths $\Gamma_k$ with $\Gamma_j = 0$ $\forall$ $j\neq k~(k=g,e,f)$.}
    \label{dephasing_dependance_fig}
\end{figure}

We extend the previous analysis with the study of the behaviour of {\it protocol 1} and {\it 2} under the effects of dephasing. While sophisticated models can be invoked to illustrate the various facets of dephasing, in order to gather an understanding of its implications for the protocols identified here, we focus on pure dephasing implemented using the 
Lindblad operators 
%
        $L_{k} = \sqrt{\Gamma_k} \op{k}{k}$,
($k \in \{e, f\}$) entering Eq.~\eqref{Lindblad1} with 
    \begin{equation}
        \mathcal{D(\rho)} = \begin{pmatrix}
                            0 & \Gamma_{ge}\rho_{ge} & \Gamma_{gf}\rho_{gf} \\
                            \Gamma_{ge}\rho_{eg} & 0 & \Gamma_{ef}\rho_{ef} \\
                            \Gamma_{gf}\rho_{fg} & \Gamma_{ef}\rho_{fe} & 0
                            \end{pmatrix},
        \label{dissipator}
    \end{equation}
where $\Gamma_{kl} =\Gamma_k + \Gamma_l$ 
 and $\rho_{kl}=\langle k|\rho|l\rangle~(k,l=g,e,f)$.
We are now able to investigate the sensitivity of the {\it protocols 1} and {\it 2} with respect to such mechanism, an analysis that we perform by independently varying the values of one of the $\Gamma_k$'s, whilst keeping the other at zero. The relationship between protocol efficiency and each of these dephasing rates can be inspected from Fig.~\ref{dephasing_dependance_fig}.
Owing to the two-photon character of the protocols at hand, we find much higher sensitivity to non-zero $\Gamma_g$ and $\Gamma_f$, whilst being comparatively resilient to $\Gamma_e$. In terms of Eq.~\eqref{dissipator}, this translates into a much larger sensitivity to $\Gamma_{gf}$ relative to $\Gamma_{ge}$, $\Gamma_{ef}$. Needless to say, this is a consequence of the protocols having been optimized to constrain the system dynamics to the subspace $\{\ket{g}, \ket{f}\}$ of the full Hilbert space and as such being most reliant on coherence between the initial and target states.

\subsection{Robustness against low-frequency noise}

\begin{figure}
{\bf (a)}\hskip3.5cm{\bf (b)}
\centering
    \includegraphics[width=\columnwidth]{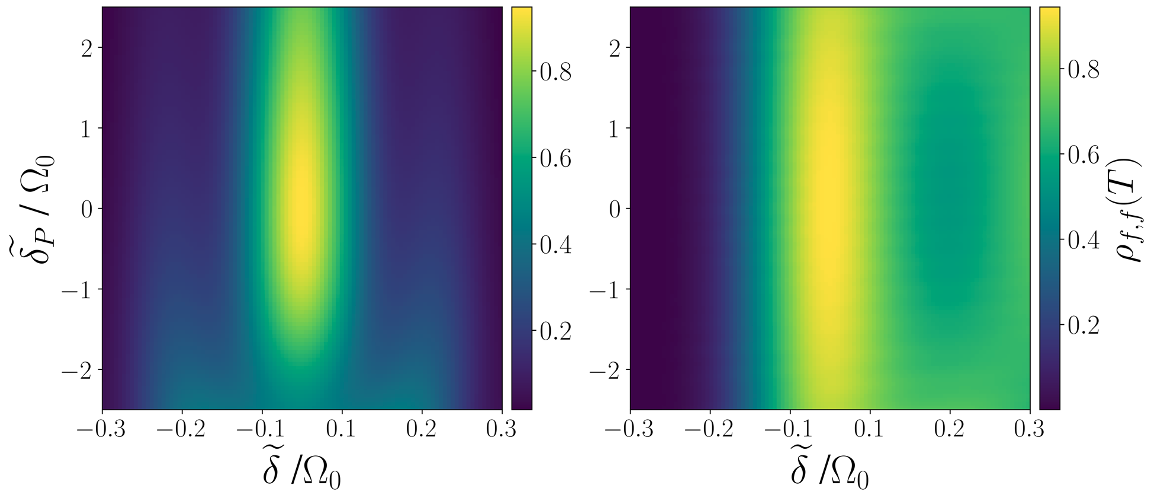}
    \caption{Sensitivity of {\bf (a)} {\it protocol 1} and {\bf (b)} {\it protocol 2} to low-frequency noise resulting in stray detunings [cf. Eq.~\eqref{stray}].}
    \label{fig:robustness}
\end{figure}

We conclude our assessment of the robustness of the proposed protocols by addressing the sensitivity to detunings. This analysis is particularly relevant for superconducting devices, where the main dephasing mechanism can be attributed to the presence of a low frequency noise that often has a $1/f$ spectrum characterized by slow fluctuations of the detunings ~\cite{elisabettaRMP,falci2013design}
. Due to the slowness of the dynamics of such fluctuations, the value of the detunings induced by such low frequency noise can be considered as constant during the population transfer. Hence, a simple way to achieve a meaningful and informative characterization of its effects on our protocols is to study their performance when we include a constant perturbation in each of the detunings. We thus take
\begin{equation}
\label{stray}
\begin{aligned}
    \delta_P (t) &\to \delta_P (t) + \tilde{\delta_P} \\
    \delta(t) &\to \delta(t) + \tilde{\delta}
\end{aligned}
\end{equation}
In the following analysis, we have considered both  such constant perturbations and the leakage mechanism outlined in Sec.~ \ref{ReinforcementLearning}. We have used the final target state population $\rho_{f,f}(T)$ as a  measure of performance. From Fig.~\ref{fig:robustness} it can be seen that both protocols are almost insensitive to  the single-photon detuning $\tilde{\delta_P}$, while 
the sensitivity to 
two-photon detuning is larger, as expected, and comparable to that of STIRAP. Interestingly, {\it protocol 2} results in $\rho_{f,f}(T)$ being strongly asymmetric with respect to the sign of the perturbation to the two photon detuning $\tilde{\delta}$. 
\FloatBarrier

\section{Conclusions}
\label{conc}
We have successfully employed a combination of a RL-based methods and more traditional optimization techniques to achieve optimal population transfer in a three-level system, whilst operating in an experimentally relevant control regime. Further, we have highlighted that our technique  can in principle be implemented as an iterative, closed-loop optimization. Its use will be beneficial in all those situations where the underlying decoherence mechanisms are not fully understood.
We have also demonstrated that even when a RL-based approach gives us sub-optimal solutions, it can still provide a useful tool that can be used to build better protocols through a simpler numerical optimization techniques.
The approach produced two novel protocols which remarkably differ from other control methods such as STIRAP, standard Raman or adiabatic schemes whilst exhibiting comparable performance and robustness. To this end, it is worth noticing that, due to the specific constraints of the protocol, STIRAP cannot be operated with both always-on couplings. 

Several works in the few years proposed the implementation of multi-level systems, including both Lambda and Ladder configurations, using superconducting artificial nanostructures subjected to suitable driving configurations. These arrangements, though, expose the nanostructure to increased noise level, which severely affects the performance of population transfer, limiting it to values that are typically in the range of $70\%$. Our approach will be invaluable to enhance the performance of such systems above and beyond the possibilities offered by demonstrated techniques for quantum control. In particular, our approach minimizes the need for the use of switchable coupling mechanisms, with is a key advantage when having in mind the design of robust schemes with low hardware overhead in the noisy intermediate-scale quantum technology framework. By serving both as an alternative control scheme for the specific physical system discussed above, and a proof-of-concept for the optimization technique itself, our protocols could be exported to be used in other relevant context, from quantum simulation to gate engineering.

\acknowledgments
This work was supported by the Northern Ireland Department for Economy (DfE), the EU H2020 framework through Collaborative Projects TEQ (Grant Agreement No. 766900),
the DfE-SFI Investigator Programme (Grant No. 15/IA/2864), the Leverhulme Trust Research Project Grant UltraQute (grant No.~RGP-2018-266), COST Action CA15220, the Royal Society Wolfson Research Fellowship scheme (RSWF\textbackslash R3\textbackslash183013) and International Mobility Programme, the UK EPSRC (grant nr.~EP/T028106/1), the Academy of Finland (
 Finnish Center of Excellence in Quantum Technology QTF projects 312296, 336810, and RADDESS programme project 328193), 
grant No.  FQXi-IAF19-06 (“Exploring the fundamental limits set by thermodynamics in the quantum regime”) of the Foundational Questions Institute Fund (FQXi), the QuantERA grant SiUCs (grant nr. 731473 QuantERA), and by University of Catania, Piano per la Ricerca 2016-18 - linea di intervento “Chance”, Piano di Incentivi per la Ricerca di Ateneo 2020/2022, proposal Q-ICT.

\appendix
\section{RL-based optimization with LSTM Neural Network}
	\label{app1}
	Traditional RL focuses on solving Markov Decision Processes (MDPs). In a MDP, the state of the environment (and the agent observation) at each time step and the corresponding action taken by the agent uniquely determine the state of the environment at the next time step~\cite{sutton1998reinforcement}.
	
	If we now consider our physical problem where the agent is trying to learn the optimal Hamiltonian of the system (under the given constraints) and the Lindblad operators are not influenced by the agent actions, the natural choice to define a MDP would be to take the density matrix of the system as the agent observation.
	
	Based on this input, we can use a function approximator (i.e. a neural network) to predict the mean values $\tilde{\mu_{\bm{\theta}}^{\delta}}$, $\tilde{\mu_{\bm{\theta}}^{\delta_P}}$ of Gaussian distributions (with standard deviation $\sigma$) whose product constitute the policy function from which we sample the actions of our agent.
	
	If a reward $R$ is granted to the agent at the end of the system dynamics, policy gradient REINFORCE~\cite{sutton1998reinforcement} can be implemented by training the neural network with a cost function $C=\frac{1}{2\sigma^2}\sum_{a_i}R|a_i-\tilde{\mu_{\bm\theta}}(s_i)|^2$, where  $\tilde{\mu_{\bm\theta}}=(\tilde{\mu_{\bm{\theta}}^{\delta}}, \tilde{\mu_{\bm{\theta}}^{\delta_P}})$ and $a_i=(\tilde{\delta}(t_i), \tilde{\delta_P}(t_i))$, with $\tilde{\delta}(t_i)$, $\tilde{\delta_P}(t_i)$ detunings normalized with respect to their maximum values (defined by their ranges).
	
	Easy improvements of this algorithm can be achived by working with a batch of agents in parallel (instead of a single agent) and training the network with stochastic gradient descent (or more advanced and similar techniques such as Adam~\cite{adam,chollet2015keras}) and subtracting a baseline to the reward (in our work we subtract the average value of the reward over the batch).
	
	This approach is effective for planning, when one can simulate the system dynamics, but it is extremely limiting as a control technique that works with real experimental data, since it requires full quantum tomography for each step of the MDP. Since measurements on a quantum system perturb the system dynamics, a good control technique would require to take measurements only at the beginning and at the end of the system evolution. Such control technique, if effective, would be more powerful than a simple application of RL to quantum systems, as it would be useful even when we are not able to simulate the system dynamics (e.g. when the noise mechanism is not fully understood).
	
	Since all the other parameters of the evolution of the system are fixed, the reward that the agent gets at the end of the process is uniquely determined by the agent actions, as the evolution of the density matrix is deterministic. Hence, in principle, the decision process in which the agent receives informations about its previous actions (that now constitute the agent observation) can be solved by means of RL techniques (and policy gradient in particular) and while defining the observation as a list of all these actions is unpractical and likely ineffective compared to other optimization techniques, we can still pursue this approach by making use of a Recurrent Neural Network (RNN) as function approximator.
	
	RNNs are neural networks specifically designed for sequential data and especially useful for time sequences. In a RNN, the output associated with each element of the input sequence depends on all the previous inputs and outputs of the network and hence this implicitly implements the desired feature. In particular we chose to use a Long Short Term Memory (LSTM) neural network that takes as external inputs only the time at which the agent is operating (details of the configuration can be found in Appendix~\ref{app2}).
	
	Comparison with standard numerical optimization techniques has been carried out in Ref.~\cite{sgroi2020reinforcement}. There, it has been shown that this approach requires a smaller number of experiments to achieve optimal protocols and shows better performances when one increases the number of control terms and the dimension of the system.

\section{\\Optimization Parameters}
\label{app2}

For the RL-based approach, we considered a batch of $N_{batch}$ agents for $N_{epochs}$ epochs of learning and we used the “Adam” optimizer ~\cite{adam,chollet2015keras} to train the Neural Network. The baseline considered is the average value of the reward over the batch. The Neural Network consists in 50 LSTM units ~\cite{chollet2015keras} followed by a dense hidden layer of 30 neurons with a Hyperbolic Tangent activation function and an output layer with the same activation function. The standard deviation of the Gaussian distribution from which the detunings are sampled is fixed to $\sigma=0.001$ for $({\delta}/{\Omega_0}$, ${\delta_P}/{\Omega_0}) \in [-50,+50]$ and to $\sigma=0.07$ for ${\delta}/{\Omega_0} \in [-0.2, +0.2]$ and ${\delta_P}/{\Omega_0} \in [-14, +14]$. The numbers of agents in the batch are, respectively $N_{batch}=100$ and $N_{batch}=50$, and the number of epochs is $N_{epochs}=350$.
The initial condition for the polynomial coefficients for the numerical optimization (Powell method) are extracted from a uniform random distribution in the interval $[-20,20]$ while for the optimization of protocol 1 and protocol 2 we used the intervals $[-5,5]$ and $[0,20]$ (for all the parameters), respectively. Throughout this work, we fixed $\Gamma T = 10$.

\FloatBarrier

\end{document}